\documentclass[vecphys]{svmult}

\usepackage{makeidx}         
\usepackage{graphicx}        
\usepackage{multicol}        
\usepackage[bottom]{footmisc}

\makeindex             

\begin{document}

\title*{Secular Evolution of Disc Galaxies and of their Components}
\author{E. Athanassoula }
\institute{LAM, Observatoire Astronomique de Marseille Provence, 2
  Place Le Verrier, \\13248 Marseille cedex 04, France. 
\texttt{lia@oamp.fr}}
%
%
\maketitle

\begin{abstract}
I discuss several aspects of secular evolution linked
to bars and to 
boxy/peanut bulges, based on a very large number of high resolution,
fully self-consistent $N$-body simulations. When the bar forms, it is
as thin as the disc. Its three-dimensional shape, however, evolves, so 
that, at later times, it has a thick inner part and a thin, more
extended outer part. The former, when viewed edge-on, is called a
boxy/peanut bulge, because of its shape. The strength of the box/peanut
correlates with the bar strength, the strongest cases having formed
after two buckling episodes. The extent of the box/peanut is considerably
shorter than the bar length, in good agreement with orbital
structure studies and with observations. Viewed at an angle near to,
but not quite edge-on, barred galaxies show specific
isodensity/isophotal shapes, which are 
different in the thick and in the thin part of the bar. The isophotes
of M31 also have such shapes. This, taken together with radial photometric
profiles and kinematics, argue that M31 is a barred galaxy. Thus, the
pseudo-ring seen at roughly 50' could be an outer ring formed at the outer
Lindblad resonance of the bar.
\end{abstract}

\section{Introduction}
\label{sec:intro}

After a short and often violent phase of formation, galaxies undergo a
long, quiet phase of evolution, called secular evolution
(e.g. Kormendy \& Kennicutt 2004).
Given its duration, this can have very important
effects on the galaxy properties. For barred galaxies, secular
evolution is driven by the bar which grows as the angular momentum
is exchanged within the galaxy. This is emitted by near-resonant
material in the bar region and absorbed by near-resonant material
in the halo and, to a much lesser extent, in the outer disc
(Athanassoula 2002, 2003). Here we will discuss specific aspects
of this secular evolution, linked to the bar and to the boxy/peanut bulge.    

\section{Box and peanut bulges}
\label{sec:peanuts}

\begin{figure}
\centering
\includegraphics[height=8cm]{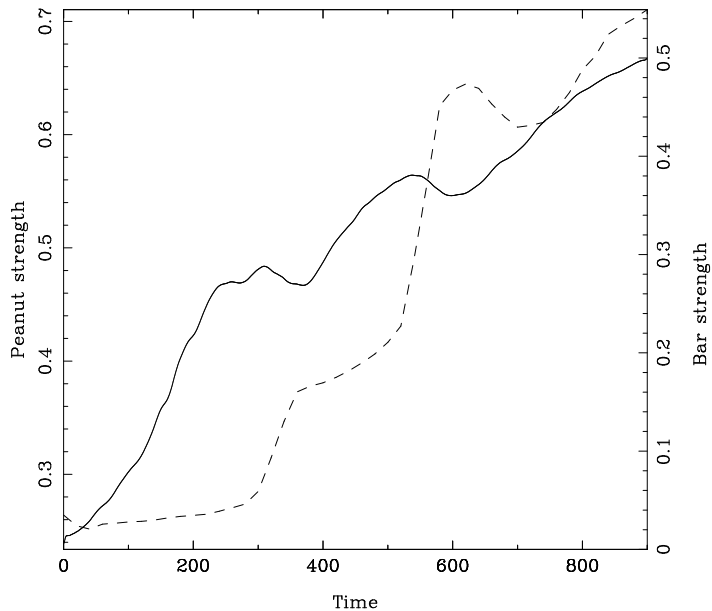}
%
%
\caption{Formation and evolution of the bar and peanut. The solid line
gives the bar strength (scale on the right) and
the dashed one the peanut strength (scale on the left), both as a
function of time for an $N$-body simulation of a disc galaxy forming a
strong bar. Time is given in computer units, with a 100 computer units
being roughly equivalent to $1.4$ Gyrs.} 
\label{fig:barpntgrowth}       
\end{figure}

\begin{figure}
\centering
\includegraphics[height=10cm]{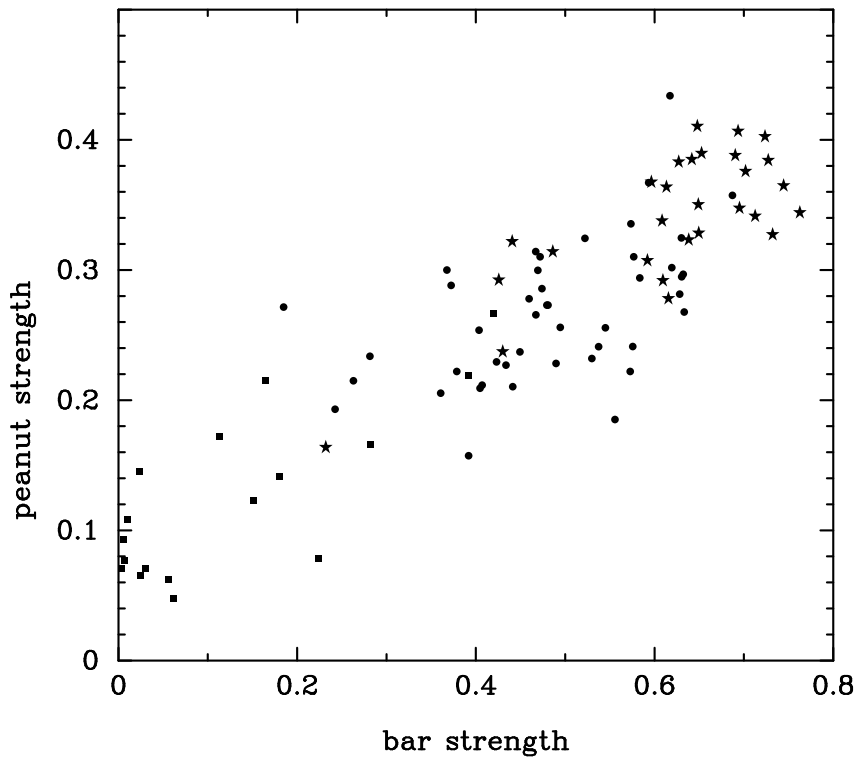}
\caption{Correlation between bar and peanut strength. Every point
  shows the result from one simulation. Simulations which underwent 0,
  1, or 2 bucklings are shown with a square, a circle, or a star, respectively.
}
\label{fig:barpntcorrel}       
\end{figure}

The backbone of two-dimensional (or very thin) bars consists of a
family of periodic orbits elongated along the bar 
and closing after one revolution around the center and two radial
oscillations (Contopoulos \& Papayannopoulos 1980; Athanassoula et al.
1983). These orbits are called x$_1$ and, when stable, they trap around them
regular orbits of roughly the same orientation and shape. The orbital
structure in three dimensions is
similar, but richer and more complicated (Pfenniger 1984; Skokos,
Patsis \& Athanassoula 2002a,b; Patsis, Skokos \& Athanassoula
2002). The backbone of three-dimensional bars is the x$_1$ tree,
i.e. the x$_1$  family plus a number of two- and three-dimensional
families bifurcating from it  
(Skokos, Patsis \& Athanassoula 2002a). The three-dimensional families,
together with the regular orbits trapped around them,
make an edge-on box or peanut shape (Patsis, Skokos \& Athanassoula
2002), such as observed in some edge-on disc galaxies and referred to
as a boxy or peanut bulge. Thus, boxy/peanut bulges are {\it
  parts} of bars seen edge-on. The interpretation of slit spectra of
such structures link them 
to bars (Kuijken \& Merrifield 1995; Bureau \&  Freeman 1999;
Merrifield \& Kuijken 1999; Chung \& Bureau 2004) as was established
by the exhaustive modeling of 
Athanassoula \& Bureau (1999) and Bureau \&
Athanassoula (2005). This link was reinforced by a study of the NIR
photometry of a sample of edge-on disc galaxies (Bureau et al. 2006). 

The formation and evolution of boxy/peanut bulges is witnessed in
many $N$-body simulations (e.g. Combes \& Sanders 1981; Combes et al. 1990;
Pfenniger \& Friedli 1991; Raha et al. 1991; Berentzen et al. 1998;
Athanassoula \& Misiriotis 2002; Athanassoula 2002, 
2003, 2005a,b,c; O'Neill \& Dubinski 2003; Debattista et al. 2004;
Martinez-Valpuesta \& Shlosman 2004; Debattista et al. 2006;
Martinez-Valpuesta, Shlosman \& Heller 2006). As shown 
Fig.~\ref{fig:barpntgrowth}, they form rather abruptly, roughly 
at the same rate as bars, but after some delay (Combes et al. 1990;
Martinez-Valpuesta \& Shlosman 2004).
The formation, often referred to as buckling, is accompanied
by a decrease of the bar strength (Raha et al. 1991;
Martinez-Valpuesta \& Shlosman 2004; Athanassoula 2005c; Martinez-Valpuesta 
et al. 2006; Debattista et al. 2006). Several cases
with more than one buckling episode have been reported in the
literature (Athanassoula 2005c; Martinez-Valpuesta et
al. 2006). Athanassoula \& Misiriotis (2002), 
based on a few simulations, showed  that the strongest bars have, when
viewed edge on, the strongest peanuts, or `X' shapes. This 
was further established, on the basis of a very large number of
simulations, by Athanassoula \& Martinez-Valpuesta (in
prep.) who find a strong correlation between bar and peanut
strength (Fig.~\ref{fig:barpntcorrel}). Thus, the trend between
bar and peanut strengths, initially found in observations
(L\"{u}tticke, Dettmar \& Pohlen 2000), is well reproduced by
simulations. Fig.~\ref{fig:barpntcorrel} also shows that simulations
in which two bucklings occurred tend to have stronger bars and peanuts
than simulations with only one buckling (Athanassoula \&
Martinez-Valpuesta, in prep.).   

\begin{figure}
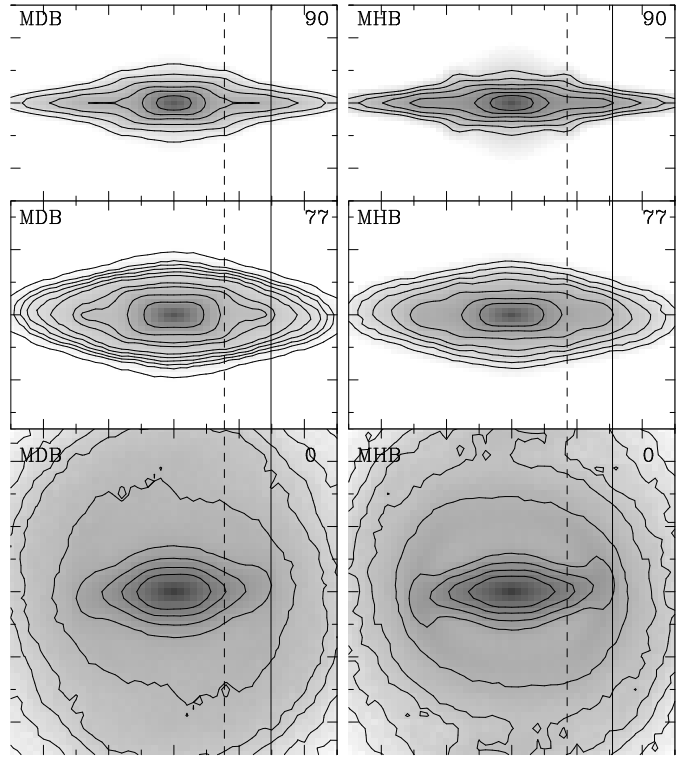

\centering
\includegraphics[height=10cm]{athanassoula_fig3a.ps}
\includegraphics[height=10cm]{athanassoula_fig3b.ps}
\caption{Three views of two simulations with a halo, a disc and a
  classical bulge. Only the two latter components are shown. In all
  cases, the bar major axis is along the $x$ axis. The upper panels
  give a side-on view (i.e. edge-on, with the line of sight along the
  bar minor axis), 
  the lower ones a face-on view and the middle ones a view at
  $77^{\circ}$. The length of the bar, as estimated from the face-on view, is
  given by a solid vertical line. The length of the box/peanut, as
  estimated by the edge-on view, is given by a vertical dashed line. It
  is clear that the extent of the boxy/peanut feature is much shorter
  than the extent of the bar. }
\label{fig:3views}       
\end{figure}

Orbital structure studies, as well as $N$-body simulations, have shown
that boxy/peanut bulges, being a {\it part} of the bar, should have a
shorter extent than that of the bar (for a discussion see Athanassoula
2005 and references 
therein). This is clearly seen in Fig.~\ref{fig:3views}, which shows the
disc and classical bulge components of two simulations at a time after
the bar and peanut have
grown. The length of the bar is estimated from the face-on view
and the length of the box/peanut from the side-on one. I extend the
vertical lines in all three panels and this makes it evident that the bar
is considerably longer than the boxy/peanut feature. Thus,  both orbital
structure studies and $N$-body simulations conclude that the bar,
being a three-dimensional object, has a thick inner part of shorter
extent and a thin outer part, of longer extent. 
This view of the bar structure is in good agreement with observations
(see Athanassoula 2005a for a compilation). Thus, it is necessary to
be careful when comparing observations to simulations, or when
comparing observations of galaxies seen at different orientations,
because a different part of the bar is seen in face-on and in edge-on
views (Fig.~\ref{fig:3views}). Further evidence that boxy/peanut
bulges are just {\it parts} of bars seen edge-on has been presented
and discussed by Athanassoula (2005a). This is based on detailed
comparisons of $N$-body simulations to observations and includes
morphology, photometry (density/light profiles along horizontal and
vertical cuts, results from median filtering) and kinematics
(cylindrical rotation, gaseous and stellar position velocity diagrams).
 
\section{M31 : A disc galaxy with a fair sized bar}
\label{sec:M31}

Fig.~\ref{fig:3views} displays also the disc and classical bulge at an
intermediate 
inclination of $77^{\circ}$, the bar major axis being again along the
galaxy major axis ($x$ axis). One can distinguish three regions with
different isodensity shapes. In the inner region, ending roughly where
the thick part of the bar ends, the isophotes have a rectangular-like
shape. This is due to the shape of the thick part of the bar. The
outermost region has elliptical isophotes, due to the disc. The
intermediate region has isophotes of a more complex shape, with two
clear protuberances (elongations) along the bar major axis on either
side of the center. This is, in fact, due to the projected shape of
the thin outer part of the bar. This intermediate region extends,  as
expected, from the end of the thick part 
of the bar to the end of the thin part (Fig.~\ref{fig:3views}).

\begin{figure}
\centering
\includegraphics[height=15cm]{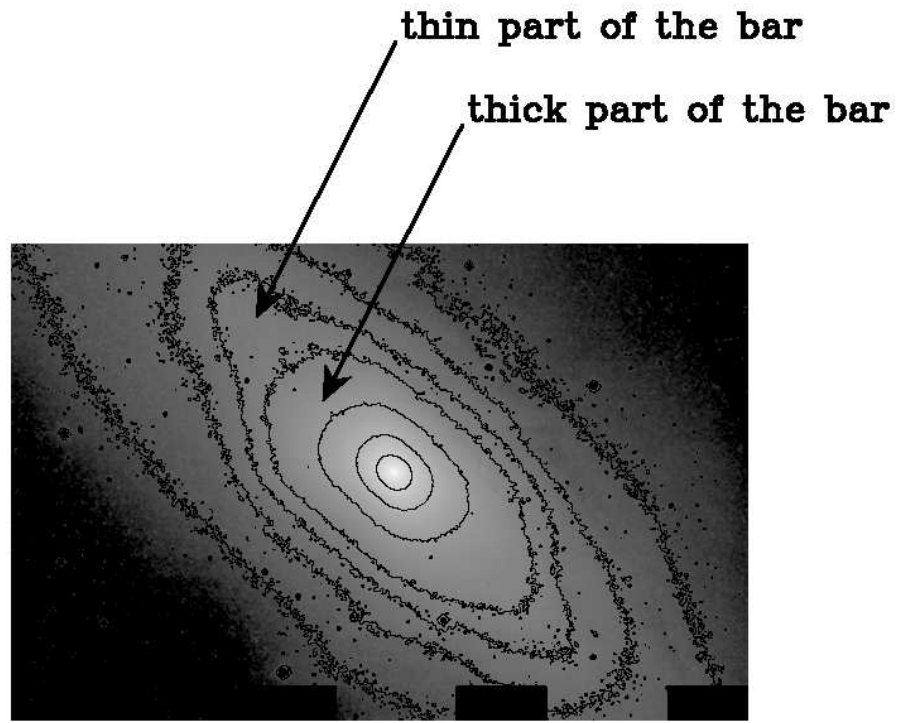}
%
%
\caption{Greyscale representation and isophotes 
    of M31 in $J$. North is at the top and East to the left.}
\label{fig:M31_iso}       
\end{figure}

Let us now compare the isodensity structure of the simulations,
discussed above, to the isophotes of M31 in the NIR, shown
in Fig.~\ref{fig:M31_iso}. Such a comparison was done initially by
Athanassoula and Beaton (2006), while the NIR data from the 2MASS
``6X survey'' are presented and discussed by Beaton et al. (2006). 
Similar structures are seen on the 3.6 $\mu$m image from the Infrared Array
Camera on the Spitzer Space Telescope (Barmby et al. 2006). It
is clear that the M31 isophotes have the same three regions as presented
above. An inner region with boxy isophotes, an outer region with
elliptical isophotes and an intermediate region with protuberances 
(elongations) pointing to a direction not far from that of the galaxy
major axis. This argues that M31 has a bar with the standard three
dimensional shape, i.e a thick inner part and a thin outer part. Such
isophotal shapes are also found in other disc galaxies observed at similar
intermediate orientations, as NGC 7582
(Quillen et al. 1997) and NGC 4442 (Bettoni \& Galletta 1994), again
revealing the existence of a bar. Athanassoula \& Beaton (2006) made a
comparison of M31 to fiducial $N$-body barred galaxy models, 
including isophotal/isodensity shapes, light/mass profiles along
cuts parallel to the galaxy major axis and kinematics. They argue that
M31 has a bar, whose length can be estimated to be roughly
22' and whose major axis is at a small angle with respect to the
galaxy major axis. They 
also present arguments that M31 has both a classical and a 
boxy/peanut bulge. This model accounts for the pseudo-ring
at roughly 50' as an outer ring due to the bar, as observed in a large
fraction of barred galaxies (e.g. Buta 1995). Indeed, 50' is
compatible with the
radius of the outer Lindblad resonance of the bar, where outer rings form
(Schwarz 1981; Athanassoula et al. 1982; Buta 1995).

A similar bar structure, with a thick inner part (consisting of the
boxy/\-pea\-nut feature) and a thin outer part of longer extent, has
been found by a number of studies of our own Galaxy (e.g. Hammersley et
al. 1994; L\`{o}pez-Corredoira 
et al. 1999; Hammersley et al. 2000; Benjamin 2005 and in
this volume; L\`{o}pez-Corredoira et al. 2006).
 
\bigskip

\noindent
{\bf Acknowledgments.} I thank A. Bosma, M. Bureau, 
I. Martinez-Valpuesta, A. Misiriotis and I. Shlosman for interesting
and motivating discussions on secular evolution and peanut
formation. This publication makes use of data products from the Two Micron
All Sky Survey, which is a joint project of the University of
Massachusetts and the Infrared Processing and Analysis
Center/California Institute of Technology, funded by the National
Aeronautics and Space Administration and the National Science
Foundation. 

%
%
%
%
%

%
%



\printindex
\end{document}